\documentclass[12pt,titlepage]{article}
\usepackage{amsfonts}
\usepackage{amsmath}
\usepackage{amssymb}

\begin{document}

\title{Comment on \\
\textquotedblleft Solutions of the Duffin-Kemmer-Petiau equation for a
pseudoscalar potential step in (1+1) dimensions\textquotedblright }
\author{T.R. Cardoso, L.B. Castro\thanks{%
benito@feg.unesp.br }, and A.S. de Castro\thanks{%
castro@pq.cnpq.br.} \\
\\
UNESP - Campus de Guaratinguet\'{a}\\
Departamento de F\'{\i}sica e Qu\'{\i}mica\\
12516-410 Guaratinguet\'{a} SP - Brazil}
\date{}
\maketitle

\begin{abstract}
It is shown that the paper \textquotedblleft Solutions of the
Duffin-Kemmer-Petiau equation for a pseu\-dosca\-lar potential step in (1+1)
dimensions\textquotedblright\ by Abdelmalek Boumali has a number of
misconceptions.
\end{abstract}

In a recent paper published in this Journal, Boumali \cite{bou} reports on
solutions of the Duffin-Kemmer-Petiau (DKP) equation and draws different
conclusions about Klein%
\'{}%
s paradox for spin-0 and spin-1 bosons. The purpose of this comment is to
point that Ref. \cite{bou} has a number of misconceptions endangering its
main conclusions.

The DKP equation for a free boson is given by \cite{pet}-\cite{duf}%
\begin{equation}
\left( i\beta ^{\mu }\partial _{\mu }-m\right) \Psi =0  \label{dkp}
\end{equation}%
\noindent where the matrices $\beta ^{\mu }$\ satisfy the algebra%
\begin{equation}
\beta ^{\mu }\beta ^{\nu }\beta ^{\lambda }+\beta ^{\lambda }\beta ^{\nu
}\beta ^{\mu }=g^{\mu \nu }\beta ^{\lambda }+g^{\lambda \nu }\beta ^{\mu }
\label{beta}
\end{equation}%
\noindent and the metric tensor is $g^{\mu \nu }=\,$diag$\,(1,-1,-1,-1)$. A
well-known conserved four-current is given by
\begin{equation}
J^{\mu }=\bar{\Psi}\beta ^{\mu }\Psi   \label{corrente}
\end{equation}%
\noindent where the adjoint spinor $\bar{\Psi}=\Psi ^{\dagger }\eta ^{0}$,
with $\eta ^{0}=2\beta ^{0}\beta ^{0}-1$ in such a way that $\left( \eta
^{0}\beta ^{\mu }\right) ^{\dagger }=\eta ^{0}\beta ^{\mu }$ (the matrices $%
\beta ^{\mu }$ are Hermitian with respect to $\eta ^{0}$). With the
introduction of interactions, the DKP equation can be written as%
\begin{equation}
\left( i\beta ^{\mu }\partial _{\mu }-m-U\right) \Psi =0  \label{dkp2}
\end{equation}%
where the more general potential matrix $U$ is written in terms of 25 (100)
linearly independent matrices pertinent to the five(ten)-dimensional
irreducible representation associated to the scalar (vector) sector. In the
presence of interactions $J^{\mu }$ satisfies the equation%
\begin{equation}
\partial _{\mu }J^{\mu }+i\bar{\Psi}\left( U-\eta ^{0}U^{\dagger }\eta
^{0}\right) \Psi =0  \label{corrent2}
\end{equation}%
Thus, if $U$ is Hermitian with respect to $\eta ^{0}$ then four-current will
be conserved. The potential matrix $U$ can be written in terms of
well-defined Lorentz structures. For the spin-0 sector there are two scalar,
two vector and two tensor terms \cite{gue}, whereas for the spin-1 sector
there are two scalar, two vector, a pseudoscalar, two pseudovector and eight
tensor terms \cite{vij}. Restricting ourselves to the spin-0 sector of the
DKP theory and considering only scalar and vector terms, $U$ is in the form%
\begin{equation}
U=S^{\left( 1\right) }+PS^{\left( 2\right) }+\beta ^{\mu }V_{\mu }^{\left(
1\right) }+i[P,\beta ^{\mu }]V_{\mu }^{\left( 2\right) }  \label{pot}
\end{equation}%
where $P$ is a projection operator ($P^{2}=P$ and $P^{\dagger }=P$) that
picks out the component of the DKP spinor which satisfies the free
Klein-Gordon equation. Note that this matrix potential leads to a conserved
four-current but the same does not happen if, instead of $i[P,\beta ^{\mu }]$
one uses $\beta ^{\mu }P$. With the representation for the $\beta ^{\mu }$\
matrices given by \cite{ned1} (apparently the same representation as that
one used in Ref. \cite{bou})%
\begin{equation}
\beta ^{0}=%
\begin{pmatrix}
\theta  & \overline{0} \\
\overline{0}^{T} & \mathbf{0}%
\end{pmatrix}%
,\quad \quad \beta ^{i}=%
\begin{pmatrix}
\widetilde{0} & \rho _{i} \\
-\rho _{i}^{T} & \mathbf{0}%
\end{pmatrix}%
,\quad i=1,2,3  \label{rep}
\end{equation}%
\noindent where%
\begin{eqnarray}
\ \theta  &=&%
\begin{pmatrix}
0 & 1 \\
1 & 0%
\end{pmatrix}%
,\quad \quad \rho _{1}=%
\begin{pmatrix}
-1 & 0 & 0 \\
0 & 0 & 0%
\end{pmatrix}
\notag \\
&&  \label{rep2} \\
\rho _{2} &=&%
\begin{pmatrix}
0 & -1 & 0 \\
0 & 0 & 0%
\end{pmatrix}%
,\quad \quad \rho _{3}=%
\begin{pmatrix}
0 & 0 & -1 \\
0 & 0 & 0%
\end{pmatrix}
\notag
\end{eqnarray}%
\noindent $\overline{0}$, $\widetilde{0}$ and $\mathbf{0}$ are 2$\times $3, 2%
$\times $2 \ and 3$\times $3 zero matrices, respectively, while the
superscript T designates matrix transposition, the projection operator can
be written as \cite{gue}%
\begin{equation}
P=\,\frac{1}{3}\left( \beta ^{\mu }\beta _{\mu }-1\right) =-\beta ^{0}\beta
^{0}\beta ^{1}\beta ^{1}\beta ^{2}\beta ^{2}\beta ^{3}\beta ^{3}=\text{diag}%
\,(1,0,0,0,0)  \label{proj}
\end{equation}%
The five-component spinor can be written as $\Psi ^{T}=\left( \Psi
_{1},...,\Psi _{5}\right) $ in such a way that the DKP equation for a boson
constrained to move along the $X$-axis decomposes into

\begin{equation*}
D_{0}^{\left( +\right) }\Psi _{1}=-i\left( m+S^{\left( 1\right) }\right)
\Psi _{2},\quad D_{1}^{\left( +\right) }\Psi _{1}=-i\left( m+S^{\left(
1\right) }\right) \Psi _{3}
\end{equation*}%
\begin{equation}
D_{0}^{\left( -\right) }\Psi _{2}-D_{1}^{\left( -\right) }\Psi _{3}=-i\left(
m+S^{\left( 1\right) }+S^{\left( 2\right) }\right) \Psi _{1}  \label{DKP3}
\end{equation}%
\begin{equation*}
\Psi _{4}=\Psi _{5}=0
\end{equation*}%
where%
\begin{equation}
D_{\mu }^{\left( \pm \right) }=\partial _{\mu }+iV_{\mu }^{\left( 1\right)
}\pm V_{\mu }^{\left( 2\right) }  \label{dzao}
\end{equation}%
In this case $J^{\mu }$ decomposes into%
\begin{equation}
J^{0}=2\text{Re}\left( \Psi _{1}\Psi _{2}^{\ast }\right) ,\quad J^{1}=-2%
\text{Re}\left( \Psi _{1}\Psi _{3}^{\ast }\right) ,\quad
J^{2}=J^{3}=0 \label{corrente3}
\end{equation}%
If the terms in the potential matrix $U$ are time-independent, one can write
$\Psi (x,t)=\psi (x)\exp (-iEt)$ in such a way that the time-independent DKP
equation decomposes into%
\begin{equation*}
\left( m+S^{\left( 1\right) }\right) \frac{d}{dx}\left( \frac{1}{m+S^{\left(
1\right) }}\frac{d\psi _{1}}{dx}\right) +2iV_{1}^{\left( 1\right) }\,\frac{%
d\psi _{1}}{dx}+k^{2}\psi _{1}=0
\end{equation*}%
\begin{equation}
\psi _{2}=\frac{1}{m+S^{\left( 1\right) }}\left( E-V_{0}^{\left( 1\right)
}+iV_{0}^{\left( 2\right) }\right) \psi _{1}  \label{dkp4}
\end{equation}%
\begin{equation*}
\psi _{3}=\frac{i}{m+S^{\left( 1\right) }}\left( \frac{d}{dx}+iV_{1}^{\left(
1\right) }+V_{1}^{\left( 2\right) }\right) \psi _{1}
\end{equation*}%
where%
\begin{eqnarray}
k^{2} &=&\left( E-V_{0}^{\left( 1\right) }\right) ^{2}-\left( V_{1}^{\left(
1\right) }\right) ^{2}+i\frac{dV_{1}^{\left( 1\right) }}{dx}+\left(
V_{0}^{\left( 2\right) }\right) ^{2}-\left( V_{1}^{\left( 2\right) }\right)
^{2}+\frac{dV_{1}^{\left( 2\right) }}{dx}  \notag \\
&&  \label{K} \\
&&-\left( m+S^{\left( 1\right) }\right) \left( m+S^{\left( 1\right)
}+S^{\left( 2\right) }\right) -\frac{iV_{1}^{\left( 1\right) }+V_{1}^{\left(
2\right) }}{m+S^{\left( 1\right) }}\,\frac{dS^{\left( 1\right) }}{dx}  \notag
\end{eqnarray}%
For this time-independent problem, $J^{\mu }$ has the components
\begin{equation}
J^{0}=2\,\frac{E-V_{0}^{\left( 1\right) }}{m+S^{\left( 1\right) }}\,|\psi
_{1}|^{2},\quad J^{1}=2\,\frac{V_{1}^{\left( 1\right) }|\psi _{1}|^{2}+\text{%
Im}\left( \frac{d\psi _{1}}{dx}\psi _{1}^{\ast }\right) }{m+S^{\left(
1\right) }}  \label{corrente4}
\end{equation}%
$J^{\mu }$ is not time dependent, so that $\psi $ describes a stationary
state.

The form $\partial _{1}+iV_{1}^{\left( 1\right) }$ in Eq. (\ref{DKP3})
suggests that the space component of the minimal vector potential can be
gauged away by defining a new spinor%
\begin{equation}
\tilde{\Psi}\left( x,t\right) =\exp \left[ i\int^{x}d\zeta \,V_{1}^{\left(
1\right) }\left( \zeta ,t\right) \right] \Psi \left( x,t\right)
\label{gauge}
\end{equation}%
even if $V_{1}^{\left( 1\right) }$ is time dependent. Furthermore, the
elimination of \ the first derivative of a second-order differential
equation, such as the term containing $V_{1}^{\left( 1\right) }$ in (\ref%
{dkp4}), is a well-known trick in mathematics. \ Nevertheless, it seems that
there is no chance to get rid from this term, except of course on condition
that one imposes $V_{1}^{\left( 2\right) }=0$.

The DKP equation is invariant under the parity operation, i.e. when $%
x\rightarrow -x$, if $V_{1}^{\left( 1\right) }$ and $V_{1}^{\left( 2\right)
} $ change sign, whereas $S^{\left( 1\right) }$, $S^{\left( 2\right) }$, $%
V_{0}^{\left( 1\right) }$ and $V_{0}^{\left( 2\right) }$ remain the same.
This is because the parity operator is $\mathcal{P}=\exp (i\delta
_{P})P_{0}\eta ^{0}$, where $\delta _{P}$ is a constant phase and $P_{0}$
changes $x$ into $-x$. Because this unitary operator anticommutes with $%
\beta ^{1}$ and $[P,\beta ^{1}]$, they change sign under a parity
transformation, whereas $\beta ^{0}$, $P$ and $[P,\beta ^{0}]$, which
commute with $\eta ^{0}$, remain the same. Since $\delta _{P}=0$ or $\delta
_{P}=\pi $, the spinor components have definite parities and the parity of $%
\Psi _{3}$ is opposite to that one of $\Psi _{1}$ and $\Psi _{2}$. The
charge-conjugation operation changes the sign of the minimal interaction
potential,\textit{\ }i.e.\textit{\ }changes the sign of \ $V_{\mu }^{\left(
1\right) }$. This can be accomplished by the transformation $\Psi
\rightarrow \Psi _{c}=\mathcal{C}\Psi =CK\Psi $, where $K$ denotes the
complex conjugation and $C$ is a unitary matrix such that $C\beta ^{\mu
}=-\beta ^{\mu }C$. The matrix that satisfies this relation is $C=\exp
\left( i\delta _{C}\right) \eta ^{0}\eta ^{1}$, where $\eta ^{1}=2\beta
^{1}\beta ^{1}+1$. The phase factor $\exp \left( i\delta _{C}\right) $ is
equal to $\pm 1$, thus $\left( \Psi _{1}\right) _{c}=$ $\mp \Psi _{1}^{\ast
} $, $\left( \Psi _{2}\right) _{c}=\pm $ $\Psi _{2}^{\ast }$, $\left( \Psi
_{3}\right) _{c}=\pm \Psi _{3}^{\ast }$ and $E\rightarrow -E$. Note also
that $J^{\mu }\rightarrow -J^{\mu }$, as should be expected for a charge
current. Meanwhile $C$ commutes with $P$ and anticommutes with $[P,\beta
^{\mu }]$, then the charge-conjugation operation entails no change on $%
S^{\left( 1\right) }$, $S^{\left( 2\right) }$ and $V_{\mu }^{\left( 2\right)
}$. By the same token it can be shown that $S^{\left( 1\right) }$and $%
S^{\left( 2\right) }$ are invariant under the time-reversal transformation
and that $V_{\mu }^{\left( 1\right) }$ and $V_{\mu }^{\left( 2\right) }$
have opposite behavior in such a way that both sorts of vector potentials
change sign under $\mathcal{PCT}$ whereas the scalar potentials do not. The
invariance of \ $S^{\left( 1\right) }$, $S^{\left( 2\right) }$ $\ $and $%
V_{\mu }^{\left( 2\right) }$ potentials under charge conjugation means that
they do not couple to the charge of the boson. In other words, $S^{\left(
1\right) }$, $S^{\left( 2\right) }$ $\ $and $V_{\mu }^{\left( 2\right) }$ do
not distinguish particles from antiparticles and the spectra for those sorts
of interactions are symmetrical about $E=0$. Hence, those sorts of
interactions can not exhibit Klein%
\'{}%
s paradox.

In summary, it is not correct to consider the most general form for the
potential matrix as being constituted by just four Lorentz structures. There
is no pseudoscalar potential in the spin-0 sector of the DKP theory. In
fact, the time component of a nonminimal vector potential is used in Ref.
\cite{bou}. It is true that $\beta ^{\mu }P$ behaves like a Lorentz vector,
nevertheless that term does not lead to a conserved current. The  operator $P
$ considered in Ref. \cite{bou} is not the proper projection operator
neither is correct the elimination $V_{1}^{\left( 2\right) }$ by a phase
transformation. Defining the transmission coefficient as the absolute value
of the ratio of the transmitted flux to the incident flux one would never
find out a negative transmission coefficient. It was a negative transmission
coefficient obtained in Ref. \cite{bou} that allowed to conclude about the
existence of Klein%
\'{}%
s paradox for a potential that does not couple to the charge of the boson.
It is also worthwhile to remark that the current expressed by (\ref{corrente}%
) is not a probability current but a charge current, always with $R+T=1$
because of the conservation law. Furthermore, there is no reason to require
that the spinor and its derivative are continuous across finite
discontinuities of the square step potential. The proper matching conditions
follow from the differential equations obeyed by the spinor components, as
they should be, avoiding in this manner the hard tasking of recurring to the
limit process of smooth potentials \cite{pla}. Finally, despite the
restriction to the one-dimensional movement, Ref. \cite{bou} treats the
problem of a particle in a (3+1)-dimensional world. In 1+1 dimensions the
matrices of the DKP algebra for spin-0 particles are reduced to 3$\times $3
matrices with three-component spinors.

\bigskip

\bigskip

\noindent

\noindent \textbf{Acknowledgments}

This work was supported in part by means of funds provided by CAPES and CNPq.

\end{document}